\documentstyle[11pt, leqno]{article}

\makeindex             

\def\rr{\bbbr}
\def\cc{\bbbc}

\def\12{\frac{1}{2}}
\def\ch{{\frak h}}
\def\chbar{\overline{\ch}}
\def\e{\E}
\def\d{{\rm d}}
\def\cH{{\cal H}}
\def\cD{{\cal D}}
\def\cO{{\cal O}}
\def\cA{{\frak A}}
\def\cB{{\cal B}}

\def\cS{{\cal S}}
\def\cW{{\cal W}}

\def\G{\Gamma}
\def\cRb{{\frak R}_{\scriptscriptstyle AW}}
\def\supp{{\rm supp\,}}
\def\wlim{{\rm w}-\lim}
\def\coinf{C_{0}^{\infty}}
\input amssym.def 
\newcommand{\beq}{\begin{equation}}
\newcommand{\eeq}{\end{equation}}
\def\EG{{^{\scriptscriptstyle E} \kern -.05cmG}}

\def\I{{I}}

\def\utw{\smash{\rlap{\lower5pt\hbox{$\sim$}}}}
\def\udtw{\smash{\rlap{\lower6pt\hbox{$\approx$}}}}

\def\bbbr{{\rm I\!R}} 

\def\bbbp{{\rm I\!P}}

\def\bbbc{{\mathchoice {\setbox0=\hbox{$\displaystyle\rm C$}\hbox{\hbox
to0pt{\kern0.4\wd0\vrule height0.9\ht0\hss}\box0}}
{\setbox0=\hbox{$\textstyle\rm C$}\hbox{\hbox
to0pt{\kern0.4\wd0\vrule height0.9\ht0\hss}\box0}}
{\setbox0=\hbox{$\scriptstyle\rm C$}\hbox{\hbox
to0pt{\kern0.4\wd0\vrule height0.9\ht0\hss}\box0}}
{\setbox0=\hbox{$\scriptscriptstyle\rm C$}\hbox{\hbox
to0pt{\kern0.4\wd0\vrule height0.9\ht0\hss}\box0}}}}
\def\bbbe{{\mathchoice {\setbox0=\hbox{\smalletextfont e}\hbox{\raise
0.1\ht0\hbox to0pt{\kern0.4\wd0\vrule width0.3pt
height0.7\ht0\hss}\box0}}
{\setbox0=\hbox{\smalletextfont e}\hbox{\raise
0.1\ht0\hbox to0pt{\kern0.4\wd0\vrule width0.3pt
height0.7\ht0\hss}\box0}}
{\setbox0=\hbox{\smallescriptfont e}\hbox{\raise
0.1\ht0\hbox to0pt{\kern0.5\wd0\vrule width0.2pt
height0.7\ht0\hss}\box0}}
{\setbox0=\hbox{\smallescriptscriptfont e}\hbox{\raise
0.1\ht0\hbox to0pt{\kern0.4\wd0\vrule width0.2pt
height0.7\ht0\hss}\box0}}}}
\def\bbbp{{\mathchoice {\setbox0=\hbox{$\displaystyle\rm P$}\hbox{\hbox
to0pt{\kern0.4\wd0\vrule height0.9\ht0\hss}\box0}}
{\setbox0=\hbox{$\textstyle\rm P$}\hbox{\hbox
to0pt{\kern0.4\wd0\vrule height0.9\ht0\hss}\box0}}
{\setbox0=\hbox{$\scriptstyle\rm P$}\hbox{\hbox
to0pt{\kern0.4\wd0\vrule height0.9\ht0\hss}\box0}}
{\setbox0=\hbox{$\scriptscriptstyle\rm P$}\hbox{\hbox
to0pt{\kern0.4\wd0\vrule height0.9\ht0\hss}\box0}}}}
\def\bbbq{{\mathchoice {\setbox0=\hbox{$\displaystyle\rm Q$}\hbox{\raise
0.15\ht0\hbox to0pt{\kern0.4\wd0\vrule height0.8\ht0\hss}\box0}}
{\setbox0=\hbox{$\textstyle\rm Q$}\hbox{\raise
0.15\ht0\hbox to0pt{\kern0.4\wd0\vrule height0.8\ht0\hss}\box0}}
{\setbox0=\hbox{$\scriptstyle\rm Q$}\hbox{\raise
0.15\ht0\hbox to0pt{\kern0.4\wd0\vrule height0.7\ht0\hss}\box0}}
{\setbox0=\hbox{$\scriptscriptstyle\rm Q$}\hbox{\raise
0.15\ht0\hbox to0pt{\kern0.4\wd0\vrule height0.7\ht0\hss}\box0}}}}
\def\bbbt{{\mathchoice {\setbox0=\hbox{$\displaystyle\rm
T$}\hbox{\hbox to0pt{\kern0.3\wd0\vrule height0.9\ht0\hss}\box0}}
{\setbox0=\hbox{$\textstyle\rm T$}\hbox{\hbox
to0pt{\kern0.3\wd0\vrule height0.9\ht0\hss}\box0}}
{\setbox0=\hbox{$\scriptstyle\rm T$}\hbox{\hbox
to0pt{\kern0.3\wd0\vrule height0.9\ht0\hss}\box0}}
{\setbox0=\hbox{$\scriptscriptstyle\rm T$}\hbox{\hbox
to0pt{\kern0.3\wd0\vrule height0.9\ht0\hss}\box0}}}}
\def\bbbs{{\mathchoice
{\setbox0=\hbox{$\displaystyle     \rm S$}\hbox{\raise0.5\ht0\hbox
to0pt{\kern0.35\wd0\vrule height0.45\ht0\hss}\hbox
to0pt{\kern0.55\wd0\vrule height0.5\ht0\hss}\box0}}
{\setbox0=\hbox{$\textstyle        \rm S$}\hbox{\raise0.5\ht0\hbox
to0pt{\kern0.35\wd0\vrule height0.45\ht0\hss}\hbox
to0pt{\kern0.55\wd0\vrule height0.5\ht0\hss}\box0}}
{\setbox0=\hbox{$\scriptstyle      \rm S$}\hbox{\raise0.5\ht0\hbox
to0pt{\kern0.35\wd0\vrule height0.45\ht0\hss}\raise0.05\ht0\hbox
to0pt{\kern0.5\wd0\vrule height0.45\ht0\hss}\box0}}
{\setbox0=\hbox{$\scriptscriptstyle\rm S$}\hbox{\raise0.5\ht0\hbox
to0pt{\kern0.4\wd0\vrule height0.45\ht0\hss}\raise0.05\ht0\hbox
to0pt{\kern0.55\wd0\vrule height0.45\ht0\hss}\box0}}}}

\def\bbbz{{\mathchoice {\hbox{$\sf\textstyle Z\kern-0.4em Z$}}
{\hbox{$\sf\textstyle Z\kern-0.4em Z$}}
{\hbox{$\sf\scriptstyle Z\kern-0.3em Z$}}
{\hbox{$\sf\scriptscriptstyle Z\kern-0.2em Z$}}}}

\def\diameter{{\ifmmode\oslash\else$\oslash$\fi}}

\newtheorem{theoreme}{Theorem }[section]

\newtheorem{proposition}[theoreme]{Proposition}

\newtheorem{lemma}[theoreme]{Lemma}
\newtheorem{definition}[theoreme]{Definition}

\newtheorem{remark}[theoreme]{Remark}

\def\rr{\bbbr}
\def\cc{\bbbc}

\def\e{{\rm e}}

\def\d{{\rm d}}
\def\12{\frac{1}{2}}

\def\proof{{\bf  Proof. }}

\def\coinf{C_{0}^{\infty}}

\def\qed{$\Box$}

\def\supp{{\rm supp\,}}
\def\cH{{\cal H}}

\def\cB{{\cal B}}
\def\G{\Gamma}

\def\cRb{{\cal R}_{\tt W}}

\def\ch{{\frak h}}

\def\cW{{\cal W}}

\def\wlim{{\rm w}-\lim}

\def\tW{W}
\def\cW{{\cal W}}

\def\cD{{\cal D}}

\def\e{{\rm e}}

\def\pfi2{P(\varphi)_{2}}

\newcommand{\ear}{\end{array}}

\begin{document}
\def\cLoc{{C^\infty_{0 \, \rr} \bigl( I^{(0)} \bigr)}}
\def\j{{\rm j}}
\def\tC{c}
\def\bc{{\rm c}}
\def\bp{{\rm p}}
\def\bt{{\rm t}}
\def\tW{{\tt W}}
\def\q{{\rm q}}
\def\chbar{\overline{\ch}}
\def\cO{{\cal O}}
\def\circle{\tt C}
\def\stp{ stochastic process }
\def\stps{stochastic processes }
\def\cUbar{\overline{\cal U}}
\def\stpos{ stochastically positive }
\def\kms{ KMS system}
\def\Xbar{\overline{X}}\def\cBbar{{\overline \cB}}
\def\cS{{\cal S}}
\def\cT{{\cal T}}
\def\cA{{\cal A}}
\def\E{{\rm e}}

\title{On the relativistic KMS condition 
for the $P(\phi)_2$ model}
\author{Christian G\'{e}rard\footnote{ \texttt{christian.gerard@math.u-psud.fr}} \\
Universit\'e Paris Sud XI,  \\
F-91405 Orsay, France\\
\and 
Christian D.\ J\"akel\footnote{\texttt{christian.jaekel@itp.phys.ethz.ch} }\\ ETH Z\"urich, H\"onggerberg\\ CH-8093 Z\"urich, Switzerland  \\ 
}
%
%
\maketitle
\abstract{The relativistic KMS condition introduced by Bros and
Buchholz provides a link between quantum statistical mechanics and
quantum field theory. 
We show that for the $P(\phi)_2$ model at positive temperature, the
two point function for fields 
satisfies the relativistic KMS condition.
}

\section{Introduction}
\label{sec:1}
The operator algebraic framework allows to characterize 
the equilibrium states of a quantum system by first principles: when
the dynamical law is changed by a local perturbation, which is slowly
switched on and slowly switched off again, an equilibrium state returns to its original form at 
the end of this procedure. 
In a pioniering work Haag, Kastler and Trych-Pohlmeyer \cite{HKT-P}
showed that this characterization of an equilibrium state leads to a sharp mathematical criterion, the 
so-called {\em KMS condition}, first encountered by Haag, Hugenholtz and Winnink~\cite{HHW} and more implicitly by 
Kubo \cite{K}, Martin and Schwinger \cite{MS}.
	
	On the other hand, the vacuum states 
of a relativistic QFT are characterized by Poincar\'e invariance and the spectrum condition \cite{SW}. Both the KMS and the spectrum condition
can be formulated in terms of analyticity properties of the correlation functions, but for almost 40 years the connection between these two
conditions was not investigated in depth, and algebraic quantum statistical mechanics and 
algebraic quantum field theory were treated in this respect as disjoint subjects.
But finally increased interest in cosmology and heavy-ion collisions led to a 
need to combine QFT and quantum statistical mechanics. 

It was first recognized by Bros and Buchholz \cite{BB1} that  the thermal equilibrium states of a relativistic QFT 
should have stronger analyticity properties in configuration space than those imposed by the traditional KMS condition. The result of their careful analysis is a {\em relativistic KMS condition} which can be understood as a remnant of (and, in applications, as a substitute for) the relativistic spectrum condition in the vacuum sector.

\subsection{Content of the Paper}

The content of this short paper is as follows. 
In Section \ref{sec:2} we recall the construction of the $P(\phi)_2$ model at positive temperature presented in \cite{GeJ}. 
In Section \ref{sec:3} we review the main ingredients of the Euclidean approach to thermal field theory. 
In Section~\ref{sec:4} we show that the Wightman two-point function
for fields satisfies the relativistic KMS 
condition. Section \ref{sec:5} is devoted to a brief discussion of the implications of the relativistic KMS condition.


\section{The spatially-cutoff $P(\phi)_2$ model at positive temperature}
\label{sec:2}

The construction of interacting thermal quantum fields in \cite{GeJ} includes several of the original ideas of H\o egh-Krohn \cite{H-K}, but instead of starting from the interacting system in a box the authors started from 
the Araki-Woods representation for the free thermal system in infinite volume. We briefly recall the main steps
of this construction.

\subsection{Preliminary Material}
\label{sec:2.1}

\paragraph{The KMS condition} 

A state $\omega_\beta$ over a $C^*$-algebra $\cA$ equipped with a
$C^{*}-$dynamics $\tau_{t}$ is called a $(\tau , \beta )$-KMS state 
for some $\beta \in \rr \cup \{ \pm \infty \}$, if for all $A, B \in \cA$  there exists a 
function $F_{A,B}$ which is continuous in the strip $\{z \in \cc \mid 0 \le \Im z \le \beta\}$ and analytic and bounded  
in the open strip $\{z \in \cc \mid 0 < \Im z < \beta\}$, with boundary values given by 
\[ F_{A, B}(t) = \omega_\beta \bigl( A \tau_t (B) \bigl),\: 
F_{A, B}(t + i\beta) = \omega_\beta \bigl( \tau_t (B) A \bigl), \:
\forall\: t\in \rr. \]

\bigskip
Thus KMS states are characterized by a real parameter $\beta$, which has the meaning of inverse temperature. 

\paragraph{The relativistic KMS condition} 

Lorentz invariance is always broken by a KMS state. A KMS state might also break spatial translation or rotation invariance, but the maximal propagation velocity of signals  is not 
 affected by such a lack of symmetry. 

\begin{definition}
A state $\omega_\beta $ over a $C^*$-algebra $\cA$ satisfies the {\em
relativistic KMS condition}    at 
inverse temperature $\beta > 0$ 
if there exists some positive timelike vector $e \in V_+$, $e^2 = 1$, such that for every pair of 
elements $A, B$ of $\cA$ there exists a function~$F_{A,B}$ which is analytic in the tube domain
\[ {\cal T}_{\beta} := \{ z \in \cc^{2} \mid \Im z \in V_+ \cap ( \beta e  + V_- ) \},  \]
and  continuous at the boundary sets $\rr^2 $ and $\rr^2 + \I \beta e $ 
with boundary values given by
\[ F_{A, B} (t,x)  = \omega_\beta \bigl(  A \alpha_{t,x} (B) \bigr)\quad 
\hbox{and}\quad
F_{A, B}  ( t+ \I \beta e, x )  = \omega_\beta \bigl( \alpha_{t,x}
(B) \bigr) \]
for all $(t,x) \in \rr^2$.
Here $V_{\pm}=\{(t,x)\in \rr^{2} \mid |t|<|x|, \: \pm t\geq 0\}$ denote the forward/backward light-cones.
\end{definition}

\begin{remark}
Note that in the limit $\beta \to \infty$ the tube ${\cal
T}_{\beta} $ tends toward the forward tube $\rr^{2} + \I V_+$.
\end{remark}

As we will discuss in Section \ref{sec:5}, the relativistic KMS condition has numerous applications. 
For instance, the famous Reeh-Schlieder property (in the thermal representation) is a direct consequence of the 
relativistic KMS condition (and weak additivity), no matter if the spatial translation or rotation invariance is broken by the KMS state or not \cite{J1}.

\subsection{The free neutral scalar field at temperature $\beta^{-1}$}
\label{sec:2.2}
Let $\ch=H^{-\12}(\rr)$ be the complex Sobolev space of order $-\12$
equipped with the norm
\[
\|h\|^{2}=(h, (2\nu)^{-1}h)_{L^{2}(\rr, \d x)}
\]for  $\nu: = (D_{x}^{2}+ m^{2})^{\12}$. 

Let  ${\frak W}(\ch)$ be the abstract Weyl $C^{*}$-algebra over $\ch$.
On ${\frak W}(\ch)$
we define  the free time evolution~$\{\tau^{\circ}_{t}\}_{t\in \rr}$
by
\[
\tau^{\circ}_{t}\bigl(W(h) \bigr)= W(\E^{\I t\nu}h), \quad h\in \ch,\: t\in \rr.
\]
For $m >0$ the unique 
$(\tau^{\circ}, \beta)$-KMS state on the Weyl algebra ${\frak W}(\ch)$ is given by
\beq
\label{dyn}
\omega^{\circ}_\beta \bigl(W(h) \bigr):= \E^{-\frac{1}{4}(h, (1+ 2\rho)h)}, \quad 
h\in \ch, 
\eeq
where $\rho:= ( \E^{\beta {\rm \nu}}-1)^{-1}$, $\beta>0$.

\paragraph{The Araki-Woods representation} 

A  realization of the  GNS representation associated to the pair~$\bigl( {\frak W}(\ch),
\omega^{\circ}_{\beta} \bigr)$ is provided by  the Araki-Woods representation.
We set:
\[
\begin{array}{l}
\cH_{\scriptscriptstyle AW}:= \G(\ch\oplus \chbar),\\[3mm]
\Omega_{\scriptscriptstyle AW}:= \Omega_{\scriptscriptstyle F} , \\[3mm]
\pi_{\scriptscriptstyle AW}(W(h))= W_{\scriptscriptstyle AW}(h):= W_{\scriptscriptstyle F} \bigl( (1+ \rho)^{\frac{1}{2}}h\oplus
\overline{\rho}^{\frac{1}{2}}\overline{h} \, \bigr),\quad h\in 
\ch.\\[3mm]
\end{array}
\] 
Here  $\overline{\ch}$ is the conjugate Hilbert space to $\ch$,
$W_{\scriptscriptstyle F}(.)$ denotes the Fock Weyl operator  on the
bosonic Fock space 
$\G(\ch\oplus\chbar)$ and $\Omega_{\scriptscriptstyle F} \in \G(\ch\oplus \chbar)$ is the Fock
vacuum.

\paragraph{The generator of the time evolution} 

Since $\omega^{\circ}_\beta$
is $\tau^\circ$-invariant, the time evolution can be unitarily implemented 
in the  representation $\pi_{\scriptscriptstyle AW}$:
\beq 
\label{Liou} \E^{i t L_{\scriptscriptstyle AW} } \pi_{\scriptscriptstyle AW} (A) \Omega_{\scriptscriptstyle AW} :=
\pi_{\scriptscriptstyle AW} \bigl(\tau_t^\circ(A)\bigr) \Omega_{\scriptscriptstyle AW}, 
\quad A \in  {\frak W}(\ch). \eeq
The generator $L_{\scriptscriptstyle AW}$ of the free time evolution,  uniquely fixed by (\ref{Liou}) and the request  $L_{\scriptscriptstyle AW} \Omega_{\scriptscriptstyle AW} =0$, is called 
the (free) Liouvillean.
It is equal to  $\d\Gamma(\nu \oplus
-\overline{\nu })$. 

\paragraph{Thermal fields} 

The Araki-Woods  representation $\pi_{\scriptscriptstyle AW}$ is a regular representation, i.e., 
the map
$\rr \ni \lambda \mapsto W_{\pi}(\lambda h)$ is strongly continuous for any $h \in \ch$.
Thus one can define {\em field operators} 
\[
\phi_{\scriptscriptstyle AW}(h):= -\I \frac{\d}{\d \lambda}W_{\scriptscriptstyle AW} (\lambda h) \Big|_{\lambda=0},\quad h \in \ch,
\]
which satisfy in the sense of quadratic forms on
$\cD(\phi_{\scriptscriptstyle AW} (h_{1}))\cap \cD(\phi_{\scriptscriptstyle AW}(h_{2}))$ the commutation
relations 
\[
[\phi_{\scriptscriptstyle AW}(h_1),\phi_{\scriptscriptstyle AW}(h_2)]=\I \Im (h_1,h_2),\quad h_1, \: h_2 \in
\ch.
\]

\paragraph{The net of local von Neumann algebras} 

The von Neumann algebra generated by $\{ \pi_{\scriptscriptstyle AW}(W(h)) \mid h \in \ch \}$
is denoted by~$\cRb$. 

To define the net of local von Neumann algebras, we introduce the
$\rr-$linear map
\[
\begin{matrix} 
{U\colon & \ch=H^{-\12}(\rr) & \to & {\cal S}'(\rr)\cr
& h & \mapsto & \Re h+ \I \nu^{-1} \Im h. \cr}
\end{matrix}
\]
For $I\subset \rr$ a bounded open interval, we define the following real vector
subspace of~$\ch$
\beq\label{defdef}
\ch_{I}:=\{h\in \ch \mid \supp Uh\subset I\}.
\eeq
We denote by $\cRb (I)$ the von Neumann 
algebra generated by
$\{W_{\scriptscriptstyle AW}(h) \mid h\in \ch_{I} \}$.  The algebra
\[
\cA:= \overline{ \bigcup_{I\subset \rr} \cRb (I)}^{(C^*)}  \]
is called the 
{\em algebra of local observables}. The union is over all open bounded
intervals  $I\subset \rr$ and
the symbol $\overline{\bigcup_{I\subset \rr} \cRb(I)}^{(C^*)}$ 
denotes the $C^{*}$-inductive limit (see e.g.~\cite[Proposition 11.4.1.]{KR}).

\begin{remark}
We note that the Araki-Woods representation is locally normal w.r.t.\ the vacuum representation of the free field.
Consequently,  the $C^*$-algebra $\cA$ is identical (up to $*$-isomorphisms) to the 
algebra 
\[
\cA_{\rm F}:= \overline{ \bigcup_{I\subset \rr} \pi_{\rm F}  \bigl(
{\frak W} (\ch_{I}) \bigr) ''}^{(C^*)},  
\]
where $\pi_{\rm F}$ is the Fock representation.
\end{remark}

\subsection{The spatially-cutoff $P(\phi)_2$ model at temperature $\beta^{-1}$} 
\label{sec:2.3}
Let $ P (\lambda) = \sum^{2n}_{j=0}a_j \lambda^j$
be a real valued polynomial, which is bounded from below, and let $l\in
\rr^{+}$ be a spatial cutoff parameter. 
The spatially cutoff $P(\phi)_{2}$  model on the real line $\rr$ at temperature
$\beta^{-1}$ is then defined  by the formal interaction term
\[
V_{l} =\int_{-l}^{l} :\! P(\phi_{AW}(x))\!: \d x, \quad l> 0.
\]
Here $: \; :$ denotes the Wick ordering (see e.g.\ \cite{GJ}) with respect to   
the {\em temperature~$\beta^{-1}$ covariance on
$\rr$}:
\beq
C_{0}(h_{1}, h_{2}):=  \Bigl( h_{1}, \frac{(1+\E^{-\beta\nu})}{
2\nu(1-\E^{- \beta \nu})}h_{2} \Bigr)_{L^{2}(\rr)},  \: \: h_{1}, h_{2}\in
\cS(\rr).
\label{c0}
\eeq
Using a sequence of functions $\{ \delta_{k} \}$ approximating the delta-function,
the limit  
\[
V_{l} := \lim_{k \to \infty} \int_{-l}^{l} :\! P(\phi_{AW}(\delta_k (.- x))\!: \d x 
\]
exists on a dense set of vectors in $\Gamma (\ch \oplus \chbar)$.
An approximation of the Dirac $\delta$ function can be fixed by setting
$\delta_{k}(x):= k \chi(kx)$, where $\chi$ is a function in $C^{\infty}_{0\:\rr} (\rr^{d})$ with $\int
\chi(x)\d x=1$.

\paragraph{The perturbed KMS system} 

It can be shown (see \cite{GeJ}) that 
the operator sum $L_{\scriptscriptstyle AW} +V_{l}$ is essentially selfadjoint on $\cD(L_{\scriptscriptstyle AW}) \cap \cD (V_{l})$ 
and  if we set
$H_{l}:=\overline{L_{\scriptscriptstyle AW} +V_{l}}$, then the perturbed time-evolution on
$\cA$ is given by 
\[ \tau^{l}_{t}(A):=\E^{\I t
H_{l}}A\E^{-\I tH_{l}} , \quad A \in \cA . \]
The perturbed KMS state $\omega_{l}$ on $\cA$ is normal w.r.t.\ the Araki-Woods 
representation~$\pi_{\scriptscriptstyle AW}$.
In fact, the GNS vector  $\Omega_{\scriptscriptstyle AW}Ê\in \Gamma(\ch\oplus\overline{\ch})$ belongs to
$\cD\bigl(\E^{-\frac{\beta}{2}H_{l}}\bigr)$ and the perturbed KMS state~$\omega_{l}$
is the vector state induced by the state vector
\[ \Omega_{l} := \frac{\E^{-\frac{\beta}{2}H_{l}}\Omega_{\scriptscriptstyle AW}}
{\|\E^{-\frac{\beta}{2}H_{l}}\Omega_{\scriptscriptstyle AW}\|} .
\]
These results are in complete analogy to the analytic perturbation theory for bounded perturbations due to Araki (see e.g.\ \cite{BR}). Identical formulas, valid for a certain class of unbounded perturbations, have recently been derived in 
\cite{DJP}.

\subsection{The translation invariant $P(\phi)_2$ model at temperature $\beta^{-1}$} 
\label{sec:2.4}

\paragraph{Existence of the  dynamics}
Let $I\subset \rr$ a bounded open interval. For~$t\in \rr$ fixed,
the norm limit
\[
\lim_{l\to \infty}\tau_{t}^{l}(A)=: \tau_{t}(A)\]
exists
for all $A\in \cRb(I)$. In fact, for $A$ and $t$ fixed,  $\tau_{t}^{l}(A)$ is independent of $l$
for $l$ sufficiently large. 

By construction the elements of the local von Neumann algebras $\cRb(I)$, 
$I$~open and bounded,  are norm dense
in~$\cA$. Thus the map $\tau \colon t \mapsto \tau_t$  
extends to a group of $*$-automorphisms of ${\frak
A}$. 
Moreover, if $\{\alpha_{x}\}_{x\in \rr}$ denotes the group of
space translations over~$\cA$, defined by
\[
\alpha_{x} \bigl(W_{\scriptscriptstyle AW}(h) \bigr):= W_{\scriptscriptstyle AW}(\E^{\I x.k}h), \:  \: x\in \rr,
\]
where $k$ is the momentum operator acting on $\ch$, then
\[ \tau_{t} \circ \alpha_{x}=\alpha_{x} \circ \tau_{t} \] 
for all $t, x\in \rr$.  Consequently the time evolution is translation invariant.

\paragraph{Existence of the thermodynamic limit}

Let $\{\omega_{l}\}_{l>0}$  be the family of $(\tau^{l},\beta)$-KMS states 
for the spatially cutoff~$P(\phi)_{2}$ models constructed in
Subsection~\ref{sec:2.3}.
Then it has been shown in \cite{GeJ} that 
\beq 
\label{thermo}
\wlim_{l\to +\infty}\omega_{l}=: \omega_\beta \hbox{ exists over }\cA.
\eeq
Moreover, the state $\omega_\beta$  has the following properties:
\vskip .3cm\halign{ \indent   #  \hfil & \vtop { 
\parindent =0pt 
\hsize=10cm                            \strut # \strut} 
\cr {\rm (i)} & $\omega_\beta$ is a $(\tau, \beta)$-KMS state over $\cA$;
\cr {\rm (ii)} &$\omega_\beta$ is {\em locally normal}, i.e., if $I \subset \rr$ is an open and bounded
interval, then $\omega_{
\beta |\cRb(I)}$ is normal 
w.r.t.~the Araki-Woods representation;
\cr {\rm (iii)} &$\omega_\beta$ is invariant under spatial translations, i.e.,
\[
\omega_\beta (\alpha_{x}(A))= \omega_\beta(A), \: \: \: x\in \rr, \: A\in \cA;
\]
\vskip -.3cm
\cr {\rm (iv)} &$\omega_\beta$ has the {\em spatial clustering property}, i.e.,
\[
\lim_{x\to\infty}\omega_\beta(A\alpha_{x}(B))= \omega_\beta(A)\omega_\beta(B) \: \: \: \: \forall  A,
B\in \cA.
\]
\vskip -.3cm
\cr}

The rate of the spatial clustering is related to the infrared properties of the Liouvillean \cite{J2}.

\section{Euclidean approach} 
\label{sec:3}
As far as the formulation of the spatially cut-off thermal $P(\phi)_2$ model is concerned, the Euclidean 
approach is only used to show that the sum of the operators $ L_{\scriptstyle AW}$ and 
$V_l$ (which are both unbounded from below) is essentially selfadjoint on the intersection of their domains.

However, for the existence of the thermodynamic limit~(\ref{thermo}) the Euclidean approach is 
used in a more sophisticated manner. The key argument in the proof of~(\ref{thermo}),  {\em Nelson symmetry}, 
will  be crucial for the proof of the relativistic KMS condition too. In order to formulate it, 
we briefly recall the Euclidean approach, in a framework adapted to
the thermal $P(\phi)_{2}$ model (see \cite{GeJ0}, \cite{KL1} for a
more general abstract framework).

\subsection{Euclidean reconstruction theorem}\label{euclrec}

\paragraph{Euclidean measures on the cylinder}

 Let $S_{\beta}=[-\beta/2, \beta/2]$ be the circle of length
$\beta$. The points in the cylinder $S_{\beta}\times \rr$ are denoted
by $(t,x)$.
Let  $Q:=\cS'_{\rr}(S_{\beta}\times \rr)$ be the space of real valued,
$\beta-$periodic 
Schwartz distributions on $S_{\beta}\times\rr$ and let $\Sigma$ be the
Borel $\sigma$-algebra on $Q$. Let $\mu$ be a Borel probability
measure on $(Q, \Sigma)$. 

For $f\in {\cal S}_{\rr}(S_{\beta}\times\rr)$, we denote by $\phi(f)$
the function
\[
\begin{matrix} 
{\phi(f)\colon &  Q & \to & \rr \cr
& q & \mapsto & \langle q, f\rangle. \cr}
\end{matrix}
\]

For $T\geq 0$, we denote by 
$\Sigma_{[0, T]}$,  the sub $\sigma$-algebra of $\Sigma$
generated by the functions $\e^{\I \phi(f)}$ for $\supp f\subset [0,
T[\times\rr$. Let  $r\colon Q\to Q$ be the time-reflection around
$t=0$ defined by $r\phi(t,x)= \phi(-t, x)$ and  let $\tau^E_{t}\colon Q\to Q$ be the
group of euclidean time translations defined by $\tau^E_{t}\phi(s, x)= \phi(s-t,x)$. 
The map $r$ lifts to a 
$*-$automorphism $R$ of $L^{\infty}(Q, \Sigma)$ defined by $RF(q):=
F(rq)$, and the group $\tau^E_{t}$ lifts to a group $U(t)$ of
$*$-automorphisms of $L^{\infty}(Q, \Sigma)$. This group is unitary on
$L^{2}(Q,\Sigma, \mu)$ if $\mu$ is invariant under $\tau^E_{t}$.

\paragraph{Reconstruction theorem}

Let $\cH_{\rm OS}:=L^{2}(Q, \Sigma_{[0, \beta/2]}, \d\mu)$. We assume
that the measure $\mu$ satisfies the {\em Osterwalder-Schrader
positivity}
\[
(F, F):=\int_{Q}R(\overline{F})F\d\mu\geq 0 \quad \forall F\in \cH_{\rm OS}.
\]
Let ${\cal N}\subset \cH_{\rm OS}$ be the kernel of $(.,.)$.  We set
\[
\cH_{\rm phys}:= \overline{\cH_{\rm OS}/{\cal N}},
\]
where the completion of $\cH_{\rm OS}/{\cal N}$ is done w.r.t. the norm $(.,.)^{\12}$. The canonical projection 
$\cH_{\rm OS}$ to $\cH_{\rm phys}$ is denoted by ${\cal V}$.

In $\cH_{\rm phys}$ we have the distinguished vector
\[
\Omega_{\rm phys}:= {\cal V}(1),
\]
where $1$ is the constant function equal to $1$ on $Q$.

The unitary group $U(t)$ for $t\geq 0$ does {\em not }preserve $\cH_{\rm
OS}$ (contrary to $0$-tem\-pe\-ra\-ture theories), because distributions
supported in the `future' $[0, \beta/2[\times \rr$ come back in the
`past' $]-\beta/2, 0]\times \rr$ by time translations. Nevertheless 
$U(s)$ for $0\leq s\leq t$ sends $L^{2}(Q, \Sigma_{[0, \beta/2-t[},
\d\mu)$ into $\cH_{\rm OS}$.

Using the theory of {\em local symmetric semigroups} (see \cite{Fr2},
\cite{KL2}), it is possible to define a selfadjoint operator
$L_{\rm phys}$ on $\cH_{\rm phys}$ such that for $F\in L^{2}(Q, \Sigma_{[0, \beta/2-t[},
\d\mu)$ and $0\leq s\leq t$ one has
\[
{\cal V}(U(s)F)= \e^{-s L_{\rm phys}}{\cal V}(F).
\]

Finally if $A\in L^{\infty}(Q, \Sigma_{\{0\}})$, then multiplication
by $A$ preserves $\cH_{\rm OS}$ and ${\cal N}$, and one obtains 
a representation $\pi_{\rm phys}$ of the {\em algebra of time-zero
fields} $L^{\infty}(Q, \Sigma_{\{0\}})$:
\[
\pi_{\rm phys}(A){\cal V}(F):= {\cal V}(AF), \quad F\in \cH_{\rm OS}.
\]
From this reconstruction procedure one obtains a $\beta$-KMS system
defined as follows:

\vskip .3cm\halign{ \indent   #  \hfil & \vtop { 
\parindent =0pt 
\hsize=10cm                            \strut # \strut} 
\cr {\rm (i)} & the $C^{*}$-algebra $\cA_{\rm phys}$ is the von Neumann algebra generated by
the operators $\e^{\I tL_{\rm phys}}\pi_{\rm phys}(A)\e^{-\I tL_{\rm
phys}}$, $t\in \rr$, $A\in L^{\infty}(Q, \Sigma_{\{0\}})$;
\cr {\rm (ii)} & the dynamics $\tau_{t}$ on $\cA_{\rm phys}$ is the dynamics generated by the
unitary group $\e^{-\I tL_{\rm phys}}$;
\cr {\rm (iii)} & the $\beta$-KMS state on $\cA_{\rm phys}$ is the state generated
by the vector $\Omega_{\rm phys}$.
\cr}

\subsection{Euclidean measure for the translation invariant  $P(\phi)_{2}$ model}
The spatially-cutoff $P(\phi)_{2}$ model at positive temperature allows an
Euclidean formulation: 
 the measure $\d\mu_{l}$ for the spatially cutoff $P(\phi)_2$ model
is given by
\beq 
\label{cutmeasure}
\d\mu_{l}:=\frac{1}{Z_l} \, \E^{-\int_{-\beta/2}^{\beta/2}\int_{-l}^{l}   : P (\phi (t,x)):_C \d t \d x}
\d\phi_{C}, \eeq 
where $\d \phi_{C}$ denotes the
Gaussian measure  on~$(Q, \Sigma)$ with
covariance 
\[ C(u, u)=(u, (D^2_t + D^2_ x + m^2)^{-1}u)\] 
(with $\beta$-periodic b.c.) defined by  
\beq
\label{e1.00}
\int_{Q}\E^{\I \phi(f)}\d\phi_{C}= \E^{- C(f,f)/2}, \quad f\in
\cS_{\rr}(S_{\beta}\times \rr).
\eeq
The partition function $ Z_l$ is chosen such that
$\int_Q d\mu_l =1$.

\paragraph {Existence of limiting measure}

In order to show that one can remove the spatial cutoff one has to show that
\beq
\lim_{l\to +\infty}\int_{Q}
F(q)\d\mu_{l}= :\int_{Q} F(q)\d\mu_\infty 
\label{limit}
\eeq
exists and defines a Borel probability measure on $\cS_{\rr}'(S_{\beta}\times\rr)$.

\paragraph {Nelson Symmetry}

Formally
exchanging the role of $t$ and $x$ in (\ref{cutmeasure})
one notices that $\d\mu_\infty $ is the Euclidean measure of the $P(\phi)_2$ 
model on the circle at temperature zero. 
This formal argument can be made rigorous (see \cite{GeJ}, \cite{H-K}).
In particular one has:
\beq
\e^{-\int_{-\beta/2}^{\beta/2}  ( \int_{-l}^{l} :  P(\phi(t, x)) :_{C_{0}}\d x ) \d t} 
= \e^{-\int_{-l}^{l} ( \int_{-\beta/2}^{\beta/2}: P(\phi(t, x)) :_{C_{\beta}}\d t ) \d x} .
\label{ns}
\eeq
Note that on the r.h.s. Wick ordering is done w.r.t. the 
covariance
\[
C_{\beta}(g_{1}, g_{2}):=\Bigl(g_{1}, \frac{1}{2\nu}g_{2}\Bigr)_{L^{2}( S_{\beta}, \d t)}, \: g_{1}, g_{2}\in
\cS(S_{\beta}).
\]
The analog of (\ref{ns}) in the zero temperature case is called {\it Nelson symmetry} (see e.g.~\cite{Si1}).

It was first noticed by H\o egh-Krohn \cite{H-K} that  the
existence of the limit~(\ref{limit}) is equivalent to the uniqueness 
of the vacuum state for the $P(\phi)_2$ model on the circle. 

Using Nelson symmetry, the  existence of the limit (\ref{limit}) is
proved in \cite{GeJ}. Moreover it is shown in \cite{GeJ} that
$\mu_{\infty}$ is OS positive, invariant under space-time
translations, and that {\em sharp-time fields} are well defined: this
means that if $\delta_{k}\in C_{0}^{\infty}(S_{\beta})$ is a sequence
of functions tending to the Dirac distribution $\delta_{0}$, then the limits
\[
\phi(t, h):=\lim_{k\to \infty}\phi(\delta_{k}(.-t)\otimes h)
\]
exist in $\bigcap_{1\leq p<\infty}L^{p}(Q, \Sigma, \d\mu_{\infty})$
for any $h\in C^{\infty}_{0}(\rr)$.
\section{The relativistic KMS condition} 
\label{sec:4}
In this section we show that two point functions for fields in the thermal $P(\phi)_2 $ model
satisfy the relativistic KMS condition.

\paragraph{Identification of physical objects}

Let $(\cH_{\beta}, \pi_{\beta}, \Omega_{\beta})$ be the GNS triple
associated to the KMS state $\omega_{\beta}$ over the $C^{*}$-algebra
$\cA$. Let also $P_{\beta}, L_{\beta}$ be the unique generators
of space-time translations such that $L_{\beta}\Omega_{\beta}=
P_{\beta}\Omega_{\beta}=0$.

As we saw in Subsect. \ref{euclrec}, the  Euclidean reconstruction
theorem  provides  a Hilbert space $\cH_{\rm phys}$, a unit
vector $\Omega_{\rm phys}$, a representation $\pi_{\rm phys}$ of the 
abelian von Neumann algebra  ${\cal U}_{\rm AW}$ generated by 
$\{W_{\rm AW}(h) \mid h\in \ch, h\hbox{ real valued }\}$, and a
selfadjoint operator $L_{\rm phys}$. Since the Euclidean measure
$\mu_{\infty}$ is  invariant under space translations, we obtain
also a selfadjoint operator $P_{\rm phys}$ implementing the space
translations.

Let us briefly check that these two families of objects are identical (up to
unitary equivalence). In the sequel we will freely identify them.

Let ${\frak U}(I)$ be the abelian von Neumann algebra generated by
$\{W_{\rm AW}(h) \mid h\in \ch, h\hbox{ real valued}, \supp h\subset
I\}$.
It was shown  in \cite{GeJ} that for $A_{j}\in {\frak U}(I)$, $1\leq
j\leq n$, 
\beq\label{toto}
\begin{array}{rl}
&(\Omega_{\rm phys}, \prod_{j=1}^{n}\e^{\I t_{j}L_{\rm phys}}\pi_{\rm
phys}(A_{j})\e^{-\I
t_{j}L_{\rm phys}}\Omega_{\rm phys})\\
=&\omega_{\beta}(\prod_{j=1}^{n}\tau_{t_{j}}(A_{j}))\\
=&(\Omega_{\beta}, \prod_{j=1}^{n}\e^{\I
t_{j}L_{\beta}}\pi_{\beta}(A_{j})\e^{-\I
t_{j}L_{\beta}}\Omega_{\beta}).
\end{array}
\eeq
Thus we can define a map $U \colon \cH_{\rm phys} \to \cH_{\beta}$ by 
\beq\label{totot}
U\e^{\I tL_{\rm phys}}\pi_{\rm phys}(A)\Omega_{\rm phys}:= \e^{\I
tL_{\beta}}\pi_{\beta}(A)\Omega_{\beta}, \quad t\in \rr, \: A\in {\frak
U}(I), \: I\subset \rr.
\eeq
From (\ref{toto}) we see that $U$ preserves the scalar product, hence
it can be uniquely extended by linearity to 
\[
{\cal E}={\rm Vect}\{\e^{\I tL_{\rm phys}}\pi_{\rm phys}(A)\Omega_{\rm phys} \mid
\: t\in \rr, A\in {\frak U}(I), \: I\subset \rr\}\]
 as an isometry.
It follows from the Euclidean reconstruction theorem, that ${\cal E}$ is dense
in $\cH_{\rm phys}$. Moreover it is clear from (\ref{totot}) that $U$ intertwines
$\pi_{\rm  phys}$ and $\pi_{\beta}$ restricted to ${\frak U}$ and
also intertwines
$L_{\rm phys}$ and $L_{\beta}$. Finally $U$ also intertwines $P_{\rm
phys}$ and $P_{\beta}$ (note that the algebra of time-zero fields is
clearly invariant under space translations).

To check that $U$ is unitary, we use
a result in \cite{GeJ}: let ${\cal B}_{\alpha}(I)$ be the von
Neumann algebra generated by $\{\tau_{t}(A) \mid  A\in {\frak U}(I),
|t|<\alpha\}$. Then it was shown in \cite[Prop. 6.5]{GeJ} that 
\[
\cRb(I)=\bigcap_{\alpha>0}{\cal B}_{\alpha}(I).
\]
Since by construction $\Omega_{\beta}$ is cyclic for
$\pi_{\beta}(\cA)$, this implies that the range of $U$ is dense
in $\cH_{\beta}$. Therefore $U$ is unitary.

\subsection{Wightman two point function for the thermal
$P(\phi)_{2}$ model}

Let $I\subset \rr$ be a bounded open interval and $\ch_{I}$ the real
subspace of $\ch$ defined in (\ref{defdef}). 

 By restriction to the local algebra ${\cal
R}_{\rm AW}(I)$, $\pi_{\beta}$ defines a CCR representation of the
real symplectic space $\ch_{I}$.
\begin{lemma}
{\it i)} The representation $\pi_{\beta}$ restricted to ${\cal
R}_{\rm AW}(I)$ is quasi-equivalent to the concrete representation of
${\cal R}_{\rm AW}(I)$;

{\it ii)} the CCR representation $\ch_{I}\ni h\mapsto \pi_{\beta}(W_{\rm
AW}(h))\in \cB(\cH_{\beta})$ is regular.
\end{lemma}
\proof 
It is well known (see e.g.\ \cite[Lemma 6.2]{GeJ}) that ${\cal R}_{\rm
AW}(I)$ is a factor. 
Now it is shown in
\cite[Prop. 10.3.14]{KR} that if ${\cal R}$ is a $C^{*}$-algebra and
$\pi$ is a factor representation of ${\cal R}$, then $\pi$ is
quasi-equivalent to the GNS representation of any $\pi$-normal state~$\omega$. 
Applying this fact to ${\cal R}_{\rm AW}(I)$ (with its
concrete representation) and to $\omega_{\beta}$, we obtain that 
$\pi_{\beta}$ is quasi-equivalent to the concrete representation
of ${\cal R}_{\rm AW}(I)$. This proves {\it i)}. We know then that there exists a $*$-isomorphism $\gamma$ from ${\cal
R}_{\rm AW}(I)$ into $\pi_{\beta}({\cal R}_{\rm AW}(I))''$ extending
$\pi_{\beta}$. This isomorphism is automatically weakly continuous.
Since the Araki-Woods representation is regular, the same holds true for
the GNS representation $\pi_{\beta}$ restricted to $\ch_{I}$. \qed

\bigskip
Since $\pi_{\beta}$ is a regular CCR representation, we can define for
$h\in \ch_{I}$ the {\em Segal field operators}
\[
\phi_{\beta}(h):=-\I\frac{\d }{\d s}\pi_{\beta}(W_{\rm
AW}(sh))\Big|_{s=0}.
\]
In the sequel we will consider only the {\em time-zero fields} $\varphi_\beta $:
\[
\varphi_{\beta}(h):= \phi_{\beta}(h), \hbox{ for }h\in \ch_{I}, \: h
\hbox{ real}.
\]

\begin{remark}
If we restrict ourselves to time-zero fields, it is possible to give a
 direct proof that the CCR representation is locally regular,  
avoiding the use of the local normality of the state
$\omega_{\beta}$. In fact, let us show that
\beq
\label{phifields}
\hbox{the map }\rr\ni s\mapsto \pi_{\beta}\bigl(W_{\rm AW}(sh)\bigr) \hbox{ is
strongly continuous for }h\in \ch_{I},\: h\hbox{ real},
\eeq
using the Euclidean approach. 

It suffices to prove the weak
continuity on a dense subspace of $\cH_{\beta}$. From the
reconstruction theorem, we see that we may take as a dense subspace of
$\cH_{\beta}$ the linear span of the vectors 
${\cal V}(\prod_{1}^{k}\e^{\I \phi(t_{j}, h_{j})})$ for $h_{j}\in
C_{0\rr}^{\infty}(\rr)$, $0\leq t_{j}<\beta/2$.

We see that it suffices to prove the continuity of the map
\[
\rr\ni s\mapsto\int_{Q} \bigl( \prod_{1}^{n}\e^{\I
\phi(t_{j}, h_{j})} \bigr) \e^{\I s \phi(0, h)}\d\mu_{\infty}
\]
for $h_{j}\in C^{\infty}_{0\rr}(\rr)$, $t_{j}\in S_{\beta}$. But this
follows from the fact that $\phi(0, h)\in L^{1}(Q, \d\mu_{\infty})$,
shown in \cite{GeJ}.
\end{remark}

\begin{lemma}
$\Omega_{\beta}\in \cD(\varphi_{\beta}(h)), \:\forall h\in\ch_{I}$, $h$
real valued.
\end{lemma}
\proof 
Clearly it suffices to prove that
\beq\label{titi}
2-(\Omega_{\beta}, \e^{\I s \varphi_{\beta}(h)}\Omega_{\beta})-
(\Omega_{\beta}, \e^{-\I s \varphi_{\beta}(h)}\Omega_{\beta})\leq
C|s|^{2}, \quad 0\leq s\leq 1.
\eeq
By the reconstruction theorem, the r.h.s.\ is equal to
\[
\int_{Q} \bigl( 2-\e^{\I s\phi(0, h)}-\e^{-\I s \phi(0, h)} \bigr) \d\mu_{\infty}.
\]
Since  $\phi(0, h)\in L^{2}(Q, \d\mu_{\infty})$, we obtain
(\ref{titi}). \qed

\bigskip

We now define Wightman two point functions. For a function $h\in
\coinf(\rr)$ we denote by $h^{-}$ the function $h^{-}(x)= h(-x)$.
\begin{proposition}
There exists a unique $\cW_{\beta}(t,.)\in C^{0}(\rr_{t}, {\cal
D}'(\rr))$ such that for $h_{1}, h_{2}\in C^{\infty}_{0 \rr}(\rr)$:
\beq\label{tutu}
\bigl(\varphi_{\beta}(h_{1})\Omega_{\beta}, \e^{\I
tL_{\beta}}\varphi_{\beta}(\alpha_{x}h_{2})\Omega_{\beta}\bigr)=
h_{1}\star h_{2}^{-}\star \cW_{\beta}(t,x), \quad (t,x)\in \rr^{2}.
\eeq
\end{proposition}
\proof For fixed $t$ the l.h.s. is a bilinear form  $Q_{t}$ w.r.t. $h_{1}$ and
$h_{2}$. Moreover it is shown in \cite{GeJ} that
$\|\varphi_{\beta}(h)\Omega_{\beta}\|\leq C\|h\|_{S}$, where
$\|.\|_{S}$ is a Schwartz seminorm. Therefore $Q_{t}$ is continuous
for the topology of $\coinf(\rr)$, which by translation invariance,
implies the existence of $\cW_\beta (t,.)$. The continuity w.r.t.\ the variable $t$ of $\cW(t,
.)$ follows from the obvious continuity in $t$ of the l.h.s. of
(\ref{tutu}). \qed

\subsection{Relativistic KMS condition}
The rest of this section is devoted to the proof of the following
theorem:
\begin{theoreme}\label{relkmsthm}
The distribution $W_{\beta}(t,x)$ extends holomorphically to 
$\rr^{2}+ \I V_{\beta}$, where $V_{\beta}:=\{(t, y) \mid \:|y|<{\rm inf}(t,
\beta-t)\}$. Therefore for $A_{i}= \varphi_{\beta}(h_{i})$, $h_{i}\in
C^{\infty}_{0\rr}(\rr)$ the two-point function $F_{A_{1}, A_{2}}(t,x)$
is holomorphic in $\rr^{2}+ \I V_{\beta}$.

\end{theoreme}
\subsection{Proof of Thm. \ref{relkmsthm}}

Let us  briefly recall a few  facts concerning the $0-$temperature
$P(\phi)_2$ model on the circle $S_\beta$. The Hilbert space is the
Fock space $\Gamma(H^{-\12}(S_{\beta}))$, where $H^{-\12}(S_{\beta})$
is the Sobolev space of order $-\12$ on $S_{\beta}$ with norm
\[
\|g\|^{2}=(g, (2b)^{-1}g)_{L^{2}(S_{\beta}, \d t)} \]
with
$b=(D_{t}^{2}+m^{2})^{\12}$. For $g\in H^{-\12}(S_{\beta})$, we denote
by $\phi_{C}(g)$ the (Fock) field operator acting on
$\Gamma(H^{-\12}(S_{\beta}))$.

The operator sum 
\[
\d\Gamma(b)+ \int_{S_{\beta}}:P(\phi_C (t)):_{C_{\beta}}\d t
\]
is essentially selfadjoint and bounded below, and the Hamiltonian of
the $P(\phi)_{2}$ model on $S_{\beta }$ is
\[
H_{C}:= \d\Gamma(b)+ \int_{S_{\beta}}:P(\phi(t)):_{C_{\beta}}\d
t-E_{C},
\]
where $E_{C}$ is an additive constant such that ${\rm
inf\ }\sigma(H_{C})=0$. The Hamiltonian $H_{C}$ has a unique (up to a
phase) ground state (i.e., vacuum state)
induced by the state vector $\Omega_{C}$. Another fact we shall need is the following: if $g_{1}, g_{2}$
are {\em real} elements of $H^{-\12}(S_{\beta})$, then
$(\phi_{C}(g_{1})\Omega_{C}, \e^{- yH_{C}}\phi_{C}(g_{2})\Omega_{C})$
is real for $y\in \rr^{+}$. This follows from the representation of
$\e^{-y H_{C}}$ using the Feynman-Kac-Nelson (FKN) formula.

Finally we note that
\[
P_{C}:=\d\Gamma(D_{t})
\]
is the momentum operator on the circle $S_\beta$. 

\paragraph{The two-point function for the $P(\phi)_{2}$ model on the circle}

Now consider the two-point function $\cW_{C}$ for the $P(\phi)_{2}$ model on
the circle
\[
 \cW_{C}(t,y)= \bigl(\Omega_{C}, \phi_{C}(\delta_{0})\e^{\I yH_{C}}\e^{\I
tP_{C}}\phi_{C}(\delta_{0})\Omega_{C}\bigr), \quad t\in S_{\beta}, \: y\in \rr
.
\]
$\cW_{C}$ is a tempered distribution on $S_{\beta}\times \rr$ rigorously
defined by
\[
\langle \cW_{C}, f\otimes g\rangle:=
(2\pi)^{\12} \bigl(\phi_{C}(\delta_{0})\Omega_{C},
\tilde{g}(-H_{C})\phi_{C}(f)\Omega_{C} \bigr),
\]
for $f\in C^{\infty}(S_{\beta})$, $g\in {\cal S}(\rr)$, and $\tilde{g}$ the Fourier transform of $g$.

To check that $\cW_{C}$ is well defined as a tempered distribution, we
use the bound
\beq\label{1}
\|(H_{C}+1)^{-\12}\phi_{C}(h)(H_{C}+ 1)^{-\12}\|\leq
C\|h\|_{H^{-1}(S_{\beta})}, 
\eeq
which using that $\delta_{0}\in H^{-1}(S_{\beta})$ yields
\[
|\langle \cW_{C} , f\otimes g\rangle|\leq C\|(H_{C}+
1)\tilde{g}(-H_{C})\|\|f\|_{H^{-1}(S_{\beta})}\|(H_{C}+
1)^{\12}\Omega_{C}\|^{2}.
\]
The r.h.s.\ can clearly be estimated in terms of Schwartz seminorms of $f$
and $g$.

\paragraph{Analytic continuation}

We first recall the {\em spectrum condition on the circle} [14]:
\[
|P_{C}|\leq H_{C}.
\]
Set 
\[
V_{\pm}:=\{(t,y)\in \rr^2 \mid |t|<|y|,\: \pm y\geq 0\}.
\]
Using $\|\e^{- \epsilon H_{C}}(H_{C}+1)\|\leq
C\epsilon^{-1}$ and the bound (\ref{1}) we conclude that 
\[
 F(\tau, z):= \bigl(\Omega_{C}, \phi_{C}(\delta_{0})\e^{\I zH_{C}}\e^{\I
\tau P_{C}}\phi_{C}(\delta_{0})\Omega_{C} \bigr)
\]
is holomorphic in $S_{\beta}\times \rr +\I V_{+}$, has a moderate
growth when $\Im (\tau, z)\to 0$ and 
\[
\begin{array}{rl}
&\int F(t+ \I \epsilon, y+ \I \epsilon) f(t)g(y)\d t \d y\\[2mm]
& \qquad =(2\pi)^{\12} \bigl(\phi_{C}(\delta_{0})\Omega_{C},
\tilde{g}(-H_{C})\e^{-\epsilon(H_{C}+ P_{C})}\phi_{C}(f)\Omega_{C} \bigr)\\[2mm]
& \qquad \to 
\langle \cW_{C}, f\otimes g\rangle,
\end{array}
\]
when $\epsilon\to 0$, i.e.,
\[
\lim_{\epsilon\to 0}F(.+ \I\epsilon, .+ \I \epsilon )= \cW_{C}(.,
.)\hbox{ in }{\cal S}'(S_{\beta}\times \rr).
\]

\paragraph{Locality on the circle}

Clearly  
\[
 \cW_{C}=\lim_{k\to +\infty}G_{k} \hbox{ in }{\cal S}'(S_{\beta}\times
\rr),
\]
where 
\[
G_{k}(t, y):= \bigl(\Omega_{C}, \phi_{C}(\delta_{k})\e^{\I yH_{C}}\e^{\I
t P_{C}}\phi_{C}(\delta_{k})\Omega_{C} \bigr),
\]
and $\delta_{k}$ is a sequence in $C^{\infty}(S_{\beta})$ with support
in $\{t \in \rr \mid |t|\leq k^{-1}\}$ and tending to~$\delta_{0}$ when $k\to \infty$.

Now using locality (i.e., finite speed of light) on the circle, we see that
if $(t,y)\in V_{\beta, k}$, 
\[
 V_{\beta, k}=\{(t,y) \mid |y|<{\rm inf}(t, \beta-t)-2k^{-1}\},
\]
then
\[
[\phi_{C}(\delta_{k}), \e^{\I yH_{C}}\e^{\I
tP_{C}}\phi_{C}(\delta_{k})\e^{-\I tP_{C}}\e^{-\I yH_{C}}]=0,
\] 
because no signal can go from ${\rm supp\ } \delta_{k}$ to ${\rm
supp\ } (\delta_{k}+ t)$ in time $y$ if $(t, y)\in V_{\beta, k}$.
This fact can be shown by exactly the same arguments as those used for
the $P(\phi)_{2}$ model on $\rr$. Thus for $(t, y)\in V_{\beta, k}$,
the function $G_{k}$ is real valued.

\paragraph{Edge of the wedge}

According to the Schwarz reflection principle, $\cW_C$ can  now be view as  the boundary value of a function
holomorphic in $V_\beta -i V_+$:
\[
 \cW_C (t, x) = H (t, x) 
\]
where
\[
 H (\tau, z) := \bigl( \phi_C (\delta_0) \Omega_C ,   \e^{- \I z H_C} \e^{- \I \tau P_C} 
\phi_C (\delta_0)   \Omega_C \bigr) . 
\]
Thus
\beq
\label{star}
\overline{\cW_C (t, y)} = \cW_C (\bar t, \bar y), \quad (t, y) \in V_\beta + i (V_+  \cup V_-). \eeq

We can now apply the edge of the wedge theorem \cite{SW}. It implies that there exists an open ball
$B (0, d):= \{ z \in \cc^2 \mid |z| < d \} $
such that $\cW_C (t,y)$ is holomorphic in $V_\beta + i \Gamma$, where 
\[ \Gamma := V_+  \cup V_- \cup B (0, d).\]

Moreover, $\cW_C(t, i y)$ is real for $t \in S_\beta$, $y > 0$ (by using the representation of
$\e^{-yH_C}$ as  a Feynman-Kac-Nelson (FKN) kernel). Thus $\cW_C (t, i y)$ is real for $t \in S_\beta$, $y \in \rr $, 
by~(\ref{star}).
Applying the Schwarz reflection principle one more time, we conclude that
\[
 \cW_C (t,y) = \cW_C (t, -y)  \quad \forall (t, y) \in V_\beta + i \Gamma. 
\]

\paragraph{Schwinger two-point function for the thermal $P(\phi)_{2}$ model}
Let $h\in C^{\infty}_{0 \rr}(\rr)$ and set
\[
I(t,x):=\int_{{\cal S}'(S_{\beta}\times \rr)}\phi(0, h)\phi(t,
\alpha_{x}h)\d\mu_{\infty}, \quad t\in S_{\beta}, x\in \rr,
\]
where $\alpha_{x}h(y)= h(y-x)$. By \cite[Thm. 7.2]{GJ}, we get:
\[
\begin{array}{rl}
I&=2\int_{-\infty<x_{1}\leq x_{2}<\infty}h(x_{1})\bigl(\Omega_{C},
\phi_{C}(\delta_{0})\e^{-(x_{2}-x_{1})H_{C}}\phi_{C}(\delta_{t})\Omega_{C}\bigr)h(x_{2}-x)\d
x_{1}\d x_{2}\\
&=2\int_{-\infty<x_{1}\leq x_{2}<\infty}h(x_{1}-x)\bigl(\Omega_{C},
\phi_{C}(\delta_{t})\e^{-(x_{2}-x_{1})H_{C}}\phi_{C}(\delta_{0})\Omega_{C}\bigr)h(x_{2})\d
x_{1}\d x_{2}.
\end{array}
\]
But $\bigl(\Omega_{C},
\phi_{C}(\delta_{0})\e^{-(x_{2}-x_{1})H_{C}}\phi_{C}(\delta_{t})\Omega_{C}\bigr)$
is real by the FKN formula and hence
\[
\bigl(\Omega_{C},
\phi_{C}(\delta_{0})\e^{-(x_{2}-x_{1})H_{C}}\phi_{C}(\delta_{t})\Omega_{C}\bigr)=\bigl(\Omega_{C},
\phi_{C}(\delta_{t})\e^{-(x_{2}-x_{1})H_{C}}\phi_{C}(\delta_{0})\Omega_{C}\bigr),
\]
which yields:
\beq
\begin{array}{rl}
I(t,x)=&2\int_{-\infty<x_{1}\leq x_{2}<\infty } \bigl( h(x_{1}) r(t, x_{2}-
x_{1}) h(x_{2}-x)
\\[3 mm]
& \qquad \qquad + h(x_{1}-x)r(t, x_{2}- x_{1}) h(x_{2})\bigr) \d x_{1}\d
x_{2},
\end{array}
\label{2}
\eeq
for
\[
 r(t, x)= \bigl(\Omega_{C},
\phi_{C}(\delta_{0})\e^{-xH_{C}}\phi_{C}(\delta_{t})\Omega_{C}\bigr)= \cW_{C}(t, \I x), \quad x\geq 0.
\]
We have seen that $\cW_{C}(t, y)= \cW_{C}(t, -y)$ for $(t,
y)\in V_{\beta}+\I \Gamma$, which when restricted to $t\in S_{\beta}$,
$\Re y=0$, yields $\cW_{C}(t, \I x)= \cW_{C}(t, -\I x)$. 

Therefore exchanging the variables $x_{1}$ and $x_{2}$ in the r.h.s. 
of (\ref{2}), we obtain:
\beq
2I(t, x)= \int_{\rr^{2}} h(x_{1})\cW_{C}(t, x_{2}- x_{1})
h(x_{2}-x)\d x_{1}\d x_{2}= h\star h^{-}\star \cW_{C}(t,x), 
\label{3}
\eeq
where $h^{-}(x):= h(-x)$. 

\paragraph{Wightman two-point function for the thermal $P(\phi)_{2}$
model}
Let $\Omega_{\beta}$ be the GNS vector for the thermal state
$\omega_{\beta}$, and let $P_{\beta}$  and $L_{\beta}$ be the generators of space-time
translations such that $P_{\beta}\Omega_{\beta}= L_{\beta}\Omega_{\beta}=0$. We  
recall that $\varphi_{\beta}(h)$ for $h\in C^{\infty}_{0\rr}(\rr)$ is the
field operator in the GNS representation.

Let $\cW_{\beta}(t,x)$ be the two-point function for the thermal
$P(\phi)_{2}$ model, defined by
\beq
\bigl(\varphi_{\beta}(h_{1})\Omega_{\beta, }\e^{\I
tL_{\beta}}\varphi_{\beta}(\alpha_{x}h_{2})\Omega_{\beta} \bigr)= h_{1}\star
h_{2}^{-}\star{\cal W}_{\beta}(t,x),
\label{5}
\eeq
for $h_{i}\in C^{\infty}_{0\rr}(\rr)$. Since $\d\mu_{\infty}$ is the
Euclidean measure for the thermal $P(\phi)_{2}$ model on the line, we
have 
\beq
{\cal W}_{\beta}(\I t, x)= {\cal W}_{C}(t, \I x), \quad 0<t<\beta/2
, \, \,  x\in \rr.
\label{4}
\eeq
As we have seen, $\cW_{C}(t, y)$ is holomorphic in
$V_{\beta}+\I \Gamma$. Thus we deduce from (\ref{4}) that $\cW_{\beta} (t,x)$ is
holomorphic in $\Gamma+\I V_{\beta}$. Applying (\ref{5}) with $h_{1}=
h_{2}=: h$, we deduce from this fact that 
\[
\varphi_{\beta}(h)\Omega_{\beta}\in {\cal D}(\e^{-(sL+yP)/2}) \quad \forall
(s, y)\in V_{\beta}.
\]
But by the spectral theorem this clearly implies that $\cW_{\beta}(t,x)$ is holomorphic in $\rr^{2}+\I V_{\beta}$.\qed

\section{Outlook}
\label{sec:5}
The condition of locality leads to strong constraints on the general form of the thermal two-point functions which allow one to apply the techniques of the Jost--Lehmann--Dyson representation. 
As has been shown by Bros and Buch\-holz~\cite{BB1}\cite{BB2}, the interacting two-point function $\cW_\beta$ can be represented in the form
\[ \cW_\beta (t,x) = \int_0^\infty {\rm d} m \, {\cal D}_\beta ( x, m ) \, {\cal W}_\beta^{(0)} ( t, x,m ). \]
Here $ {\cal D}_\beta ( x, m ) $ is a distribution in $ x, m$ which is symmetric in $x$, and 
\[Ê{\cal W}_\beta^{(0)} (x,m) = (2 \pi)^{-1}  \int \d \nu \d p  \,\varepsilon (\nu) \delta (\nu^2 - p^2 - m^2) (1 - \E^{- \beta \nu} )^{-1} \E^{i (\nu t - px)}  \]
is the two-point correlation function of the free field of mass $m$ in a thermal equilibrium state at inverse temperature $\beta$. In contrast to the vacuum case, the damping factors ${\cal D}_\beta (x, m ) $ depend in general in a non-trivial way on the spatial variables $x$. They describe the dissipative effects of the thermal background on the propagation of sharply localized excitations.  

If the underlying equilibrium state satisfies the relativistic KMS condition, the function $D_\beta (x, m )$ 
is regular in $x$ and admits an analytic continuation into the domain 
$ \{ z \in \cc  \mid | \Im  z  | < \beta / 2 \} $.

\end{document}